\documentclass[10pt]{article}

\usepackage{authblk}

\usepackage[english]{babel} 
\usepackage{graphicx}
\usepackage{wallpaper}
\usepackage{amsmath}
\usepackage{amsfonts}
\usepackage{amsthm}
\usepackage{amscd}
\usepackage{amssymb}
\usepackage{bm}
\usepackage{subfigure}

\usepackage{graphicx}
\usepackage{epstopdf}
\usepackage{dcolumn}
\usepackage{bm}
\usepackage{color}
\usepackage{amssymb, amsmath, bm, textcomp}

\usepackage[T1]{fontenc} 
\usepackage{float}       
\usepackage{mathrsfs}

\usepackage{color}
\usepackage{nicefrac}
\usepackage{lineno}

\newcommand{\be}{\begin{equation}}
\newcommand{\ee}{\end{equation}}
\newcommand{\bea}{\begin{eqnarray}}
\newcommand{\eea}{\end{eqnarray}}
\usepackage[a4paper,top=4.5cm,left=3cm,right=3cm,bottom=2cm]{geometry}

\begin{document}

\begin{center}

{\LARGE
\textbf\newline{\textbf{Simulating Soft-Sphere Margination \\ in Arterioles and Venules}}
}

\vspace*{0.5cm}

Giacomo Falcucci$^{1,2,*,+}$,  Simone Melchionna$^{3,+}$, Paolo Decuzzi$^{4,+}$ and Sauro Succi$^{2,5,+}$

\vspace*{0.5cm}

\begin{small}

\textit{$^1$Dept. of Enterprise Engineering ``Mario Lucertini'' - University of Rome ``Tor Vergata'' \\
Via del Politecnico 1, 00133, Roma (Italy) \\
$^2$John A. Paulson School of Engineering and Applied Sciences - Harvard University \\
33 Oxford Street, 02138 Cambridge (USA) \\
$^3$Istituto Sistemi Complessi - National Research Council, c/o Dept. of Physics - University of Rome ``Sapienza'' \\
Piazzale Aldo Moro, 00100, Rome (Italy) \\
$^4$Laboratory of Nanotechnology for Precision Medicine - Istituto Italiano di Tecnologia \\
Via Morego 30, 16163 Genova (Italy) \\
$^5$Istituto per le Applicazioni del Calcolo - National Research Council, Via dei Taurini 19, 00185 Roma (Italy)}

\textit{$^*$giacomo.falcucci@uniroma2.it}

\textit{$^+$these authors contributed equally to this work}
\end{small}
\end{center}


\begin{abstract}
In this paper, we deploy a Lattice Boltzmann - Particle Dynamics (LBPD) method to dissect the transport properties within arterioles and venules.
First, the numerical approach is applied to study the transport of Red Blood Cells (RBC) through plasma and validated by means of comparison with the experimental data in the seminal work by F\aa hr\ae us and Lindqvist. \\
Then, the presence of micro-scale, soft spheres within the blood flow is considered: the evolution in time of the position of such spheres is studied, in order to highlight the presence of possible \textit{margination} effects. \\
The results of the simulations and the evaluation of the computational effort shows that LBPD offers a computational appealing strategy for the study of complex biological flows.

\end{abstract}


%
\thispagestyle{empty}

\section{Introduction}

Since the seminal work by Matsumura et al., \cite{matsumura1986new}, in which the enhanced permeability and retention (EPR) effect of blood vessel wall was found in tumoral tissues, the adoption of micro- and nano-particles for drug delivery has been made the object of remarkable research efforts, \cite{blanco2015principles}.
The presence of abnormal fenestration in the blood vessels of injured tissues, as well as the internal lesions due to plaque exfoliation and thrombosis provide the opportunity for local increased accumulation of long-circulating particles through mechanisms such as adhesion and extravasation, \cite{matsumura1986new,maeda2016retrospective,maeda2017polymer}.

Today, despite the remarkable efforts and progress in the last 30 years, only about $1 \%$ of injected micro- and nano-particles reach the target area, \cite{Mura2013,anchordoquy2017mechanisms}, due to the heterogeneity and to the limited accessibility of biological target cells.
Among the reasons for such a limited targeting, it is known that the fluid dynamic pattern within blood vessels plays a major role on the distribution of carriers: in particular, the size, shape, surface properties and mechanical stiffness (abbreviated as the ``4S Parameters'', \cite{anchordoquy2017mechanisms}) are known to have a paramount relevance in the vascular transport of micro- and nano-particles, together with the geometric layout of the blood vessel, with its bifurcations and remarkable variations in section, \cite{decuzzi2010size}.

In recent years, the constant increase in computational power has elicited the flourishing of numerical simulations of blood flow, in order to dissect the complex aspects related to the presence of red blood cells (RBC's), platelets and other organic and inorganic compounds dispersed in the plasma liquid phase, \cite{bernaschi_petaflop_2011,melchionna_model_2011,fedosov_predicting_2011,reasor2013determination,mountrakis2014validation}

Particular attention is being conveyed to the \textit{margination} of transported micro- and nano-particles. Margination within the blood flow consists in the migration of suspended particles or cells toward vessel walls: it has been observed experimentally for white blood cells, platelets and rigid micro-particles, but the dependence of margination on the ``4S Parameters'', \cite{anchordoquy2017mechanisms}, as well as on vessel geometry  \cite{decuzzi2010size}, remains an open challenge, so far, \cite{mountrakis2013platelets,muller2014margination}.

In this paper, we employ the LBPD Method to dissect the aspects related to blood transport in small vessels.
LBM has been successfully employed in recent years for brad-range phenomena of scientific and technical interest, \cite{Falcucci_Book_2018,Falcucci2007,FalcucciSM2010,Falcucci2011a,succi_MRT,Falcucci_Harvard,falcucci2017heterogeneous} across physical scales, and it has proven to be a reliable and versatile tool for very large-scale simulation of biological phenomena, \cite{bernaschi_flexible_2010,melchionna_model_2011,Succi_2009}. \\
First, we present a validation of MUPHY code \cite{melchionna_muphy_2009} by comparing the relative apparent viscosity obtained through numerical simulations to the experimental results by  F\aa hr\ae us and Lindqvist (FL), \cite{faahraeus1931viscosity}. To the best of the authors' knowledge, this is the first time that the Lattice Boltzmann Method is used to model the FL effect in a wide range of vessel diameters. 
Then, for two values of vessel section, we study the transport of micro-scale, non-deformable spheres, characterized by a diameter of 4 $\mu$m, the same as the smaller dimension of RBC's.
The locations of such micro-spheres are randomly initialized at the beginning of the simulation ($t=0$) and their \textit{margination} behavior along the simulation is studied, considering three value of sphere concentration.

The results provide a reasonable behavior of micro-scale transported particles, compared to experimental and numerical works in the literature; moreover, the proposed procedure results in a very light computational effort, thus emphasizing the reliability and versatility of the proposed approach for complex multi-scale transport phenomena within the blood flow.

\section{Results and Discussion}

\subsection{Model Validation: the F\aa hr\ae us-Lindqvist Effect}

\begin{figure*}
\begin{center}
	\includegraphics[angle=-90,width=0.79\textwidth]{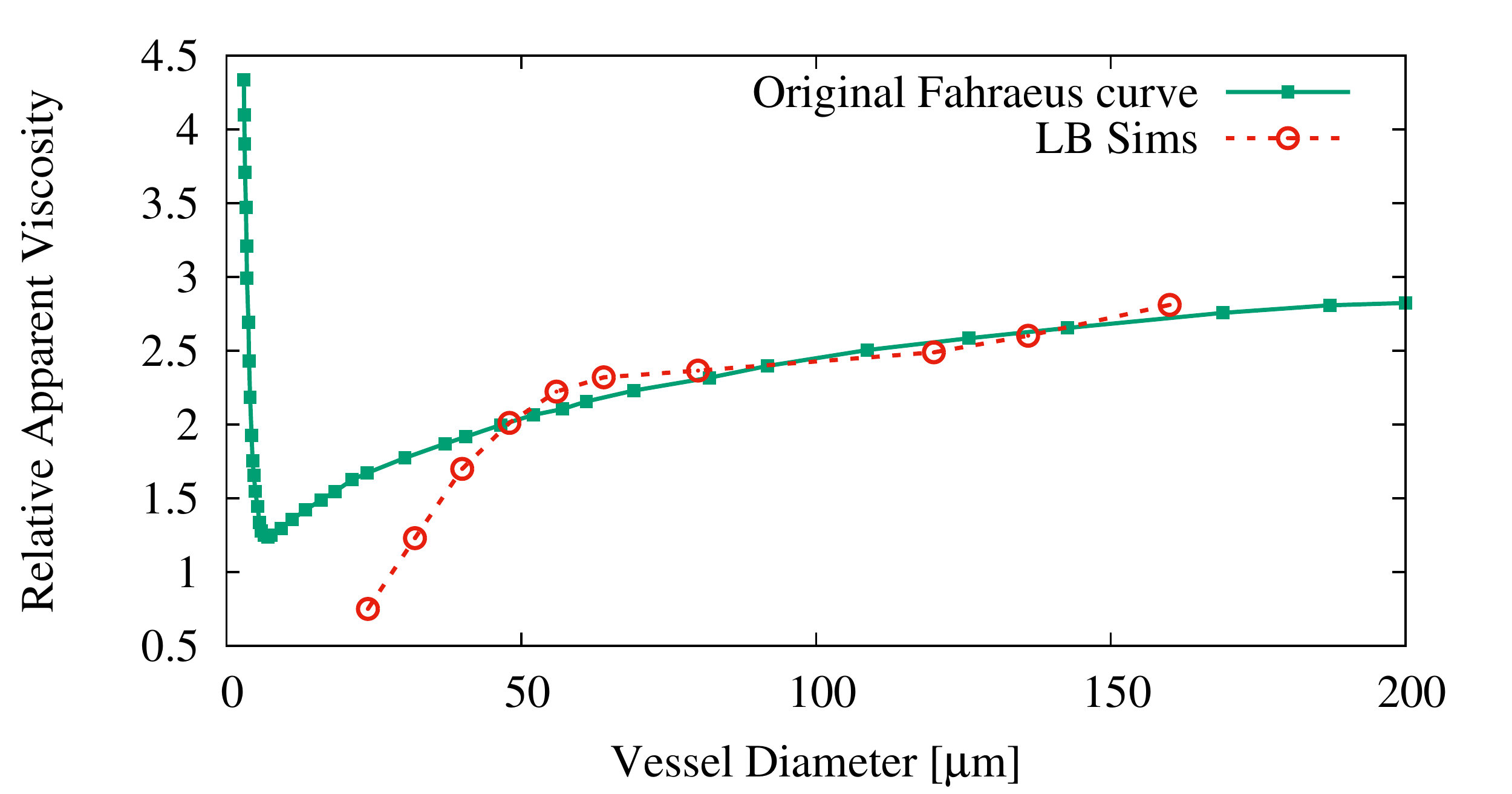}
	\caption{\label{Fig_Fahr} Trend of Relative apparent viscosity as a function of the vessel diameter. The solid line represents the experimental curve found by F\aa hr\ae us and Lindqvist, \cite{faahraeus1931viscosity}; the dotted line reports the results of our LB simulations. The minimum in the experimental trend is located at a vessel diameter of $\sim 8 \ \mu$m, corresponding to the size of RBC's.}
\end{center}
\end{figure*}

Figure \ref{Fig_Fahr} reports the validation of the proposed methodology by means of comparison with F\aa hr\ae us-Lindqvist (FL) curve, \cite{faahraeus1931viscosity}.
According to this seminal work, in fact, it is known that the presence of Red Blood Cells (RBC's) dispersed within the plasma phase produces a characteristic \textit{non-Newtonian} effect: the mixture viscosity varies with the diameter of the blood vessel, as highlighted by the green line in Fig. \ref{Fig_Fahr}, \cite{faahraeus1931viscosity,vahidkhah2016flow}.
The relative apparent viscosity is computed by comparing the volumetric flow rate at different vessel diameters with an hematocrit value of $HCT = 40 \%$ with the corresponding Poiseuille flow characterized by the flow of pure plasma. Thus, we can write:
\be
 		\nu_{rel} = \frac{Q_{Poiseuille}}{Q_{HCT=40 \%}}, 
\ee
in which $Q_{HCT=40 \%} = S_{vessel} \left[ u_{plasm} (1 - HCT) + u_{RBC} \ HCT \right]$; $S_{vessel}$ is the area of the cross section of the vessel; $u_{plasm}$ is the velocity of the plasm phase; $u_{RBC}$ is the velocity of the erythrocytes.
The value of $\nu_{rel}$ is supposed to be always larger than unity, $\nu_{rel} > 1$, as the characteristic viscosity of RBC's is larger than that of the plasma phase, \cite{fedosov_multiscale_2010}. Moreover, the FL effect is enhanced by the presence of a cell free layer (CFL) close to the channel walls, with the erythrocytes moving towards the center of the vessel, \cite{haynes1960physical,mcwhirter2009flow,fedosov2010blood,fedosov2010multiscale,reasor2013determination,mountrakis2014validation}. 

Figure \ref{Fig_Fahr} shows the very good agreement between the experimental data and the numerical simulation in the range $40 \ \mu \text{m} < D_{vessel} < 160  \ \mu \text{m}$, in presence of a consistent value of $HCT$. \\
The Figure highlights that the  accuracy of our numerical predictions deteriorates as the vessel diameter decreases.
At diameters smaller than $40 \ \mu \text{m}$, in fact, the effects related to RBC deformability start to play a dominant role, as the channel section reduces towards the dimensions of the erythrocytes, characterized by a biconcave disc shape with $8 \ \mu$m $\times 8 \ \mu$m $\times 4 \ \mu$m dimensions: to be noted, the minimum value of $\nu_{rel}$ in the FL trend is located at capillary diameters equal to the RBC larger section.
Moreover, the interactions with compliant walls become more and more important at smaller sections: future works will be aimed at optimizing MUPHY to account for these effects.

\subsection{Blood Cell Rheology}

In the range of good agreement between experiments and simulations, we chose two values of vessel diameter, namely $D_1=80 \ \mu$m and $D_2=120 \ \mu$m.
For these cases, we consider the presence of soft-potential, non-deformable, micro-spheres dispersed within the blood flow. 
It is important to stress that in the present simulations, such micro-spheres are randomly dispersed within the computational domain at the beginning of the simulation: no ``master-profiles'' along the vessel diameter are considered for the sphere location, \cite{reasor2013determination,karnadakis_communication}. \\
Table \ref{simcases} reports the cases considered in the following.

\begin{table}[h!]
   \begin{center}
\begin{tabular}{l c c c }
\hline
\hline
																										&   $N_{RBC}$			&	$N_{SPH}$ 			& $D_{SPH}$ \\	
\hline	
$D_{vessel} = 80 \mu$m, $N_{SPH} =0.5 \% \ N_{RBC}$	\qquad &		3485					&			17			&	$4.0 \ \mu$m	 \\
$D_{vessel} = 80 \mu$m, $N_{SPH} =1 \% \ N_{RBC}$		\qquad &		3485					&			34			&	$4.0 \ \mu$m	 \\
$D_{vessel} = 80 \mu$m, $N_{SPH} =3 \% \ N_{RBC}$		\qquad &		3485					&			104		&	$4.0 \ \mu$m	 \\
\hline
$D_{vessel} = 120 \mu$m, $N_{SPH} =0.5 \% \ N_{RBC}$	\qquad &		11948				&			59			&	$4.0 \ \mu$m	 \\
$D_{vessel} = 120 \mu$m, $N_{SPH} =1 \% \ N_{RBC}$		\qquad &		11948				&			119		&	$4.0 \ \mu$m	 \\
$D_{vessel} = 120 \mu$m, $N_{SPH} =3 \% \ N_{RBC}$		\qquad &		11948				&			358		&	$4.0 \ \mu$m	 \\			
\hline
\hline
\end{tabular}
\caption{\label{simcases} Main physical parameters for the simulations with micro-spheres. The number of RBC's corresponds to an hematocrit $HCT = 40 \%$.}
\end{center}
\end{table}

\begin{figure*}
	\subfigure[$D=80 \ \mu$m]{\includegraphics[width=\textwidth]{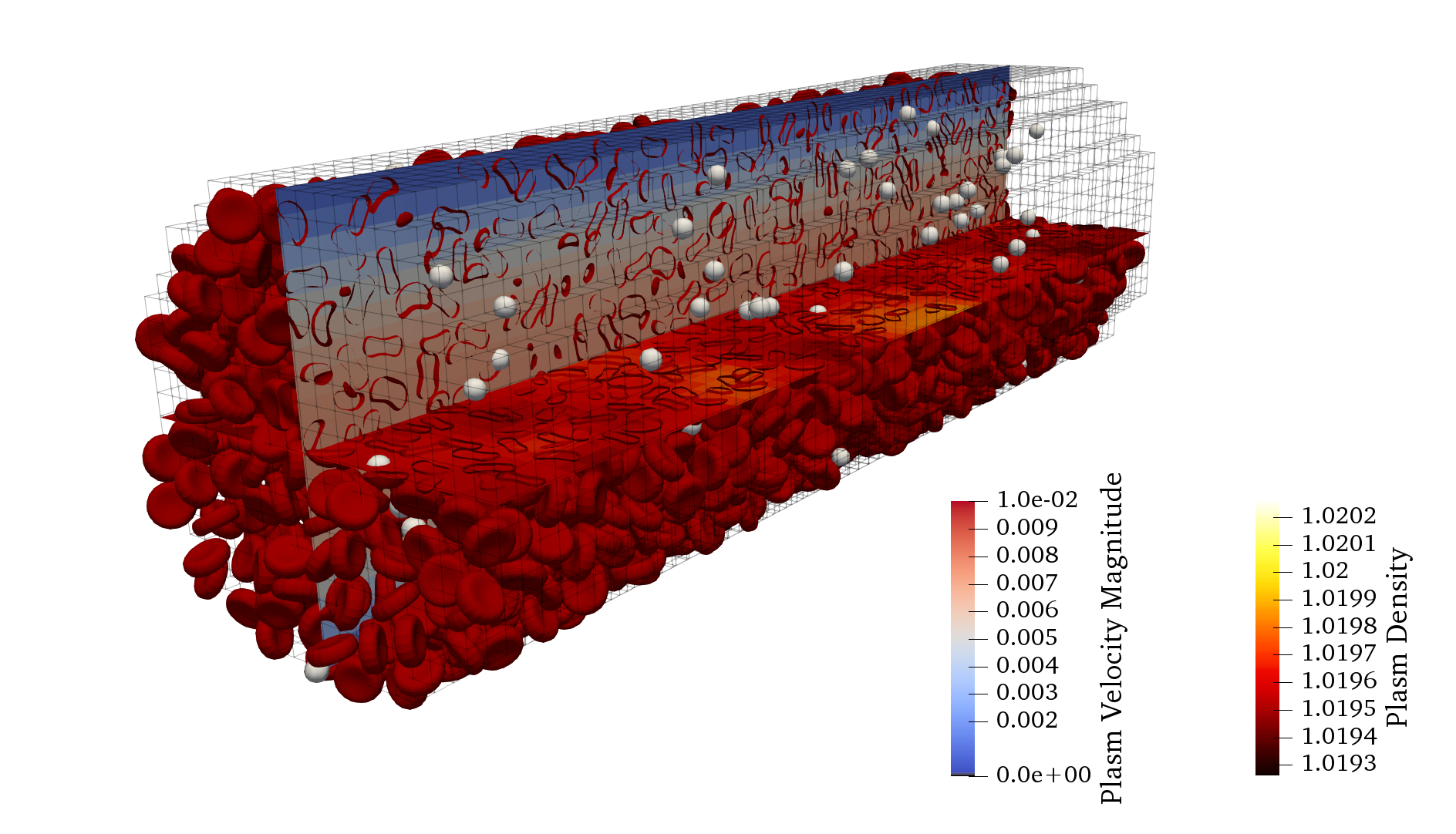}} \\
	\subfigure[$D=120 \ \mu$m]{\includegraphics[width=\textwidth]{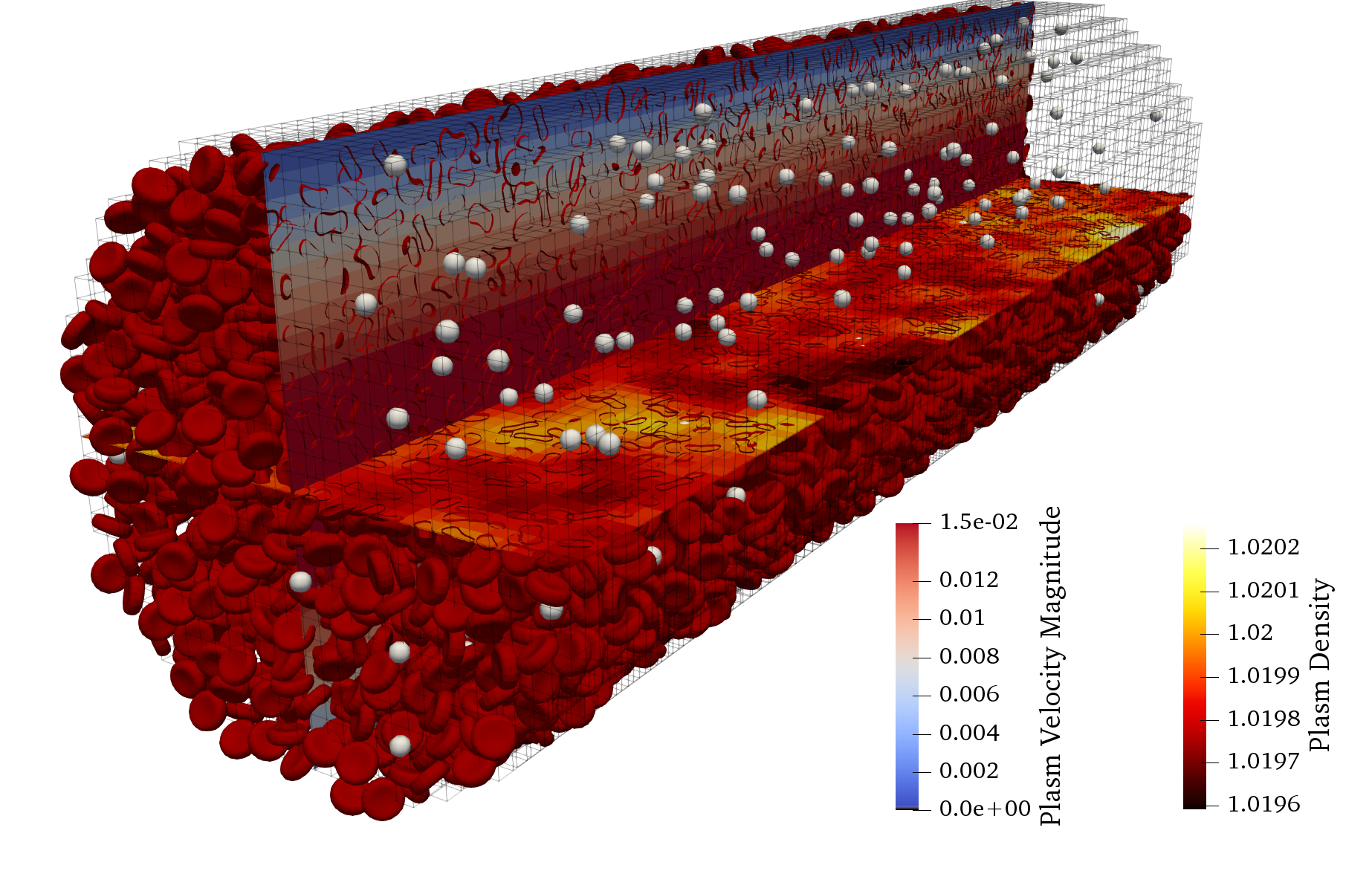}}
    \caption{\label{Vess_Sect} Section of the blood vessels corresponding to $R=40 \ \mu$m and $R=60 \ \mu$m. Details of RBC's and Spheres are reported. The cross sections show the velocity magnitude and density fields of the plasm phase.}
\end{figure*}

\begin{figure}
	\begin{center}
			\subfigure[$R= 40 \mu$m]{\includegraphics[angle=-90,width=0.52\textwidth]{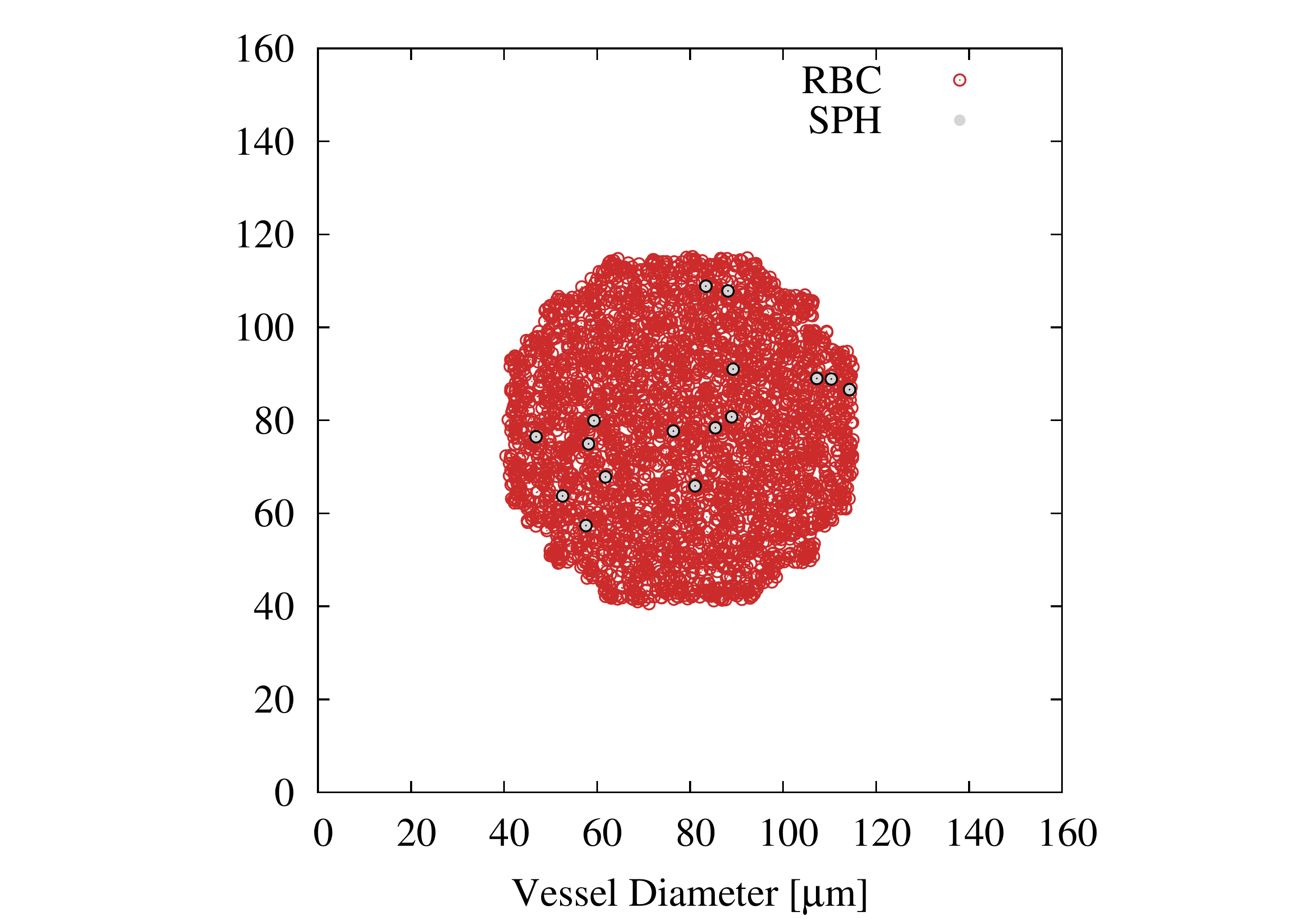}} \hspace{-1.cm}
			\subfigure[$R= 60 \mu$m]{\includegraphics[angle=-90,width=0.52\textwidth]{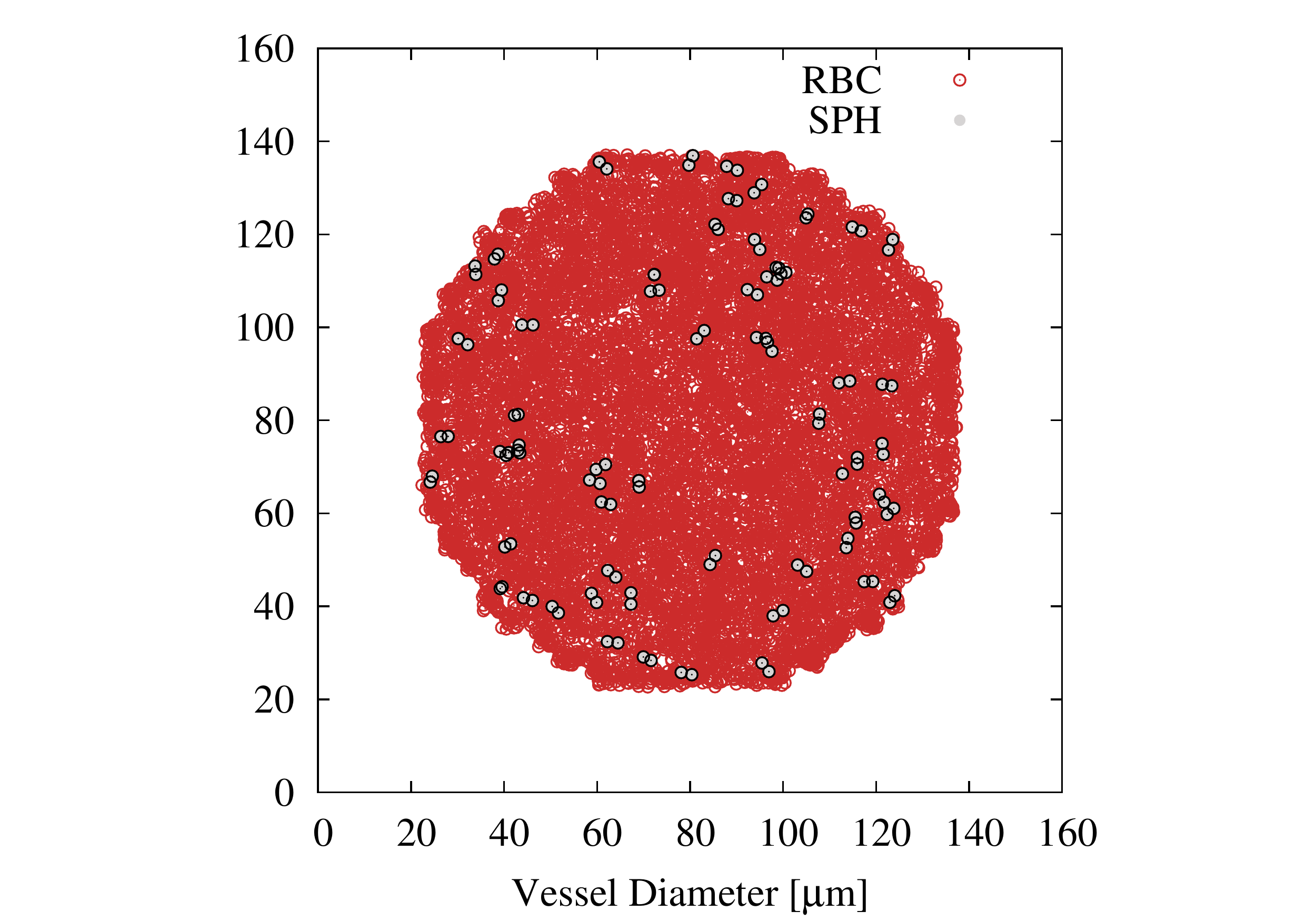}}
			\caption{\label{Fig_R_60_mum_05} (a) Statistical distribution of RBC and SPH inside the vessel for $R=40 \ \mu$m and $R=60 \ \mu$m; (b) cross section of the vessel, with the erythrocytes and spheres reported at $t > 10^6$ time steps, corresponding to steady state flow conditions. The presence of the cell free layer is apparent from the bar chart and confirmed by the right channel section. In the panel, the number of spheres corresponds to $0.5 \%$ of the total number of erythrocytes.}
	\end{center}
\end{figure}

Panel \ref{Fig_R_60_mum_05} shows the statistical distribution of RBC's and Spheres for the case corresponding to 
$HCT= 40 \%$, for the two chosen vessel dimensions. The Figure reports the locations of RBC's and Spheres in the cross section of the channel for the case corresponding to $N_{SPH} = 0.5 \ \% \ N_{RBC}$.
Despite the random initialization of the spheres, Figure \ref{Vess_Sect} suggests an appreciable \textit{margination} effect towards the outer of the channel.

\begin{figure}
	\begin{center}
			\subfigure[$N_{SPH} = 0.5 \% N_{RBC}$]{\includegraphics[angle=-90,width=0.34\textwidth]{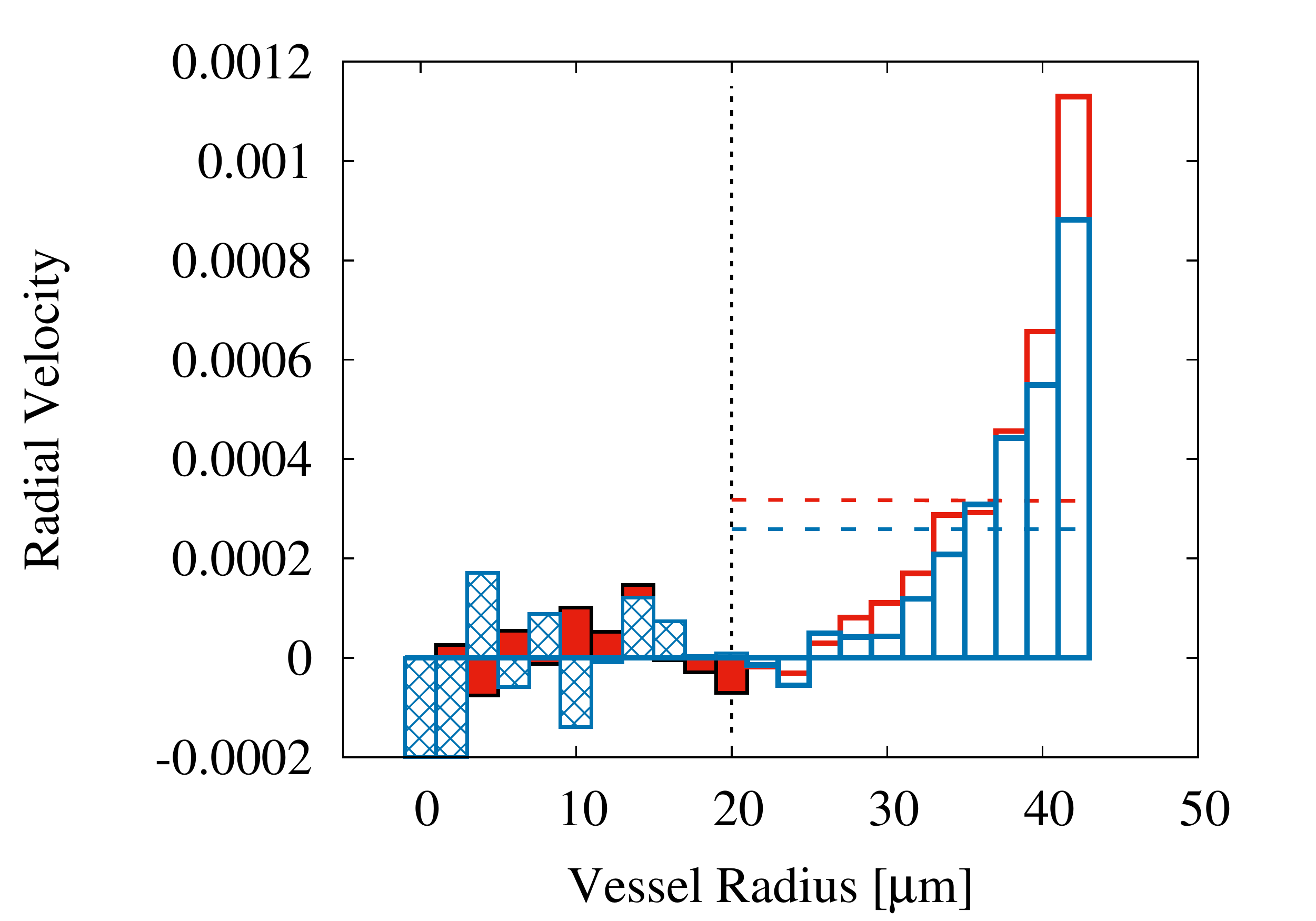}} \hspace{-0.5cm}
			\subfigure[$N_{SPH} = 1 \% N_{RBC}$]{\includegraphics[angle=-90,width=0.34\textwidth]{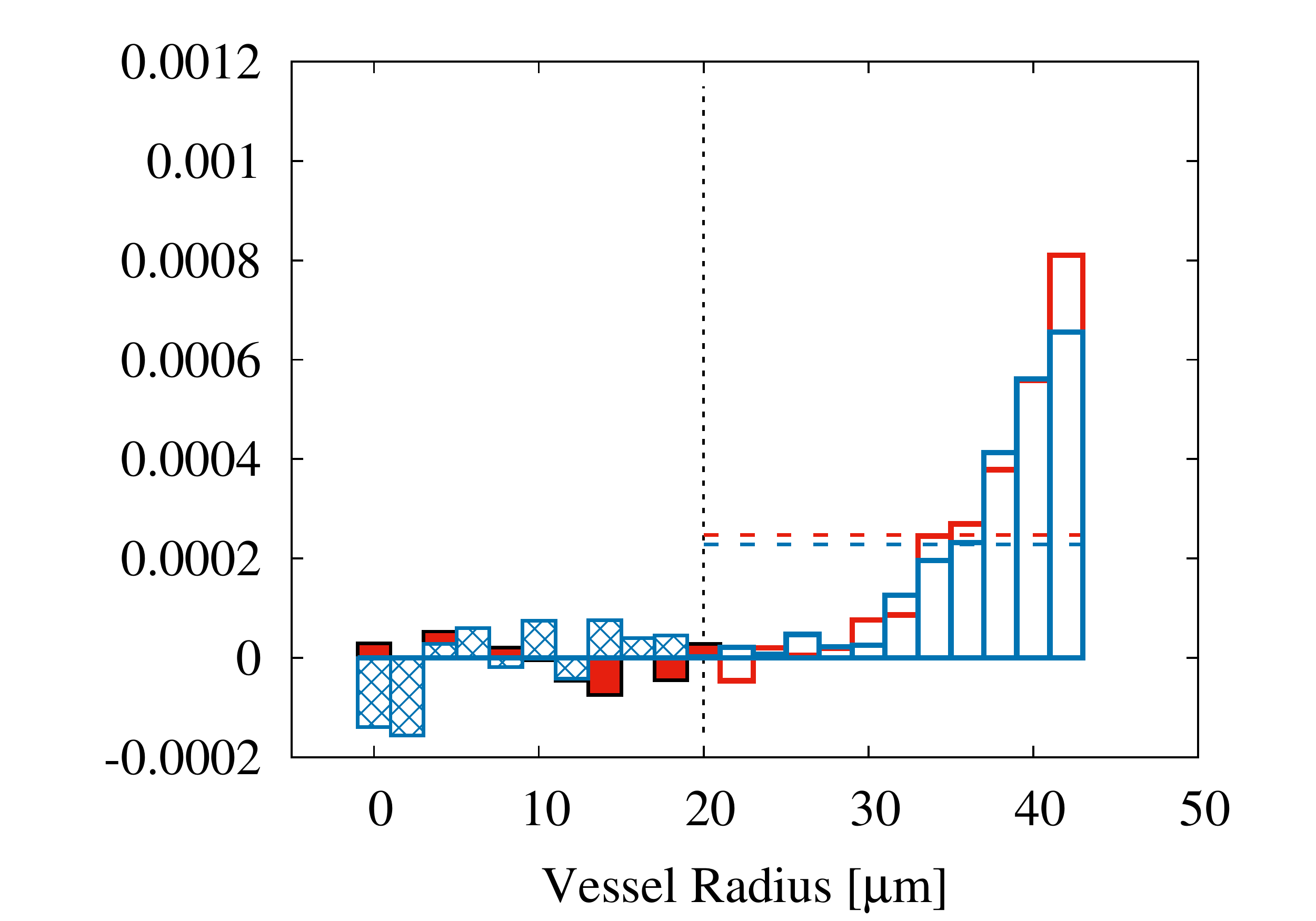}} \hspace{-0.5cm}
			\subfigure[$N_{SPH} = 3 \% N_{RBC}$]{\includegraphics[angle=-90,width=0.34\textwidth]{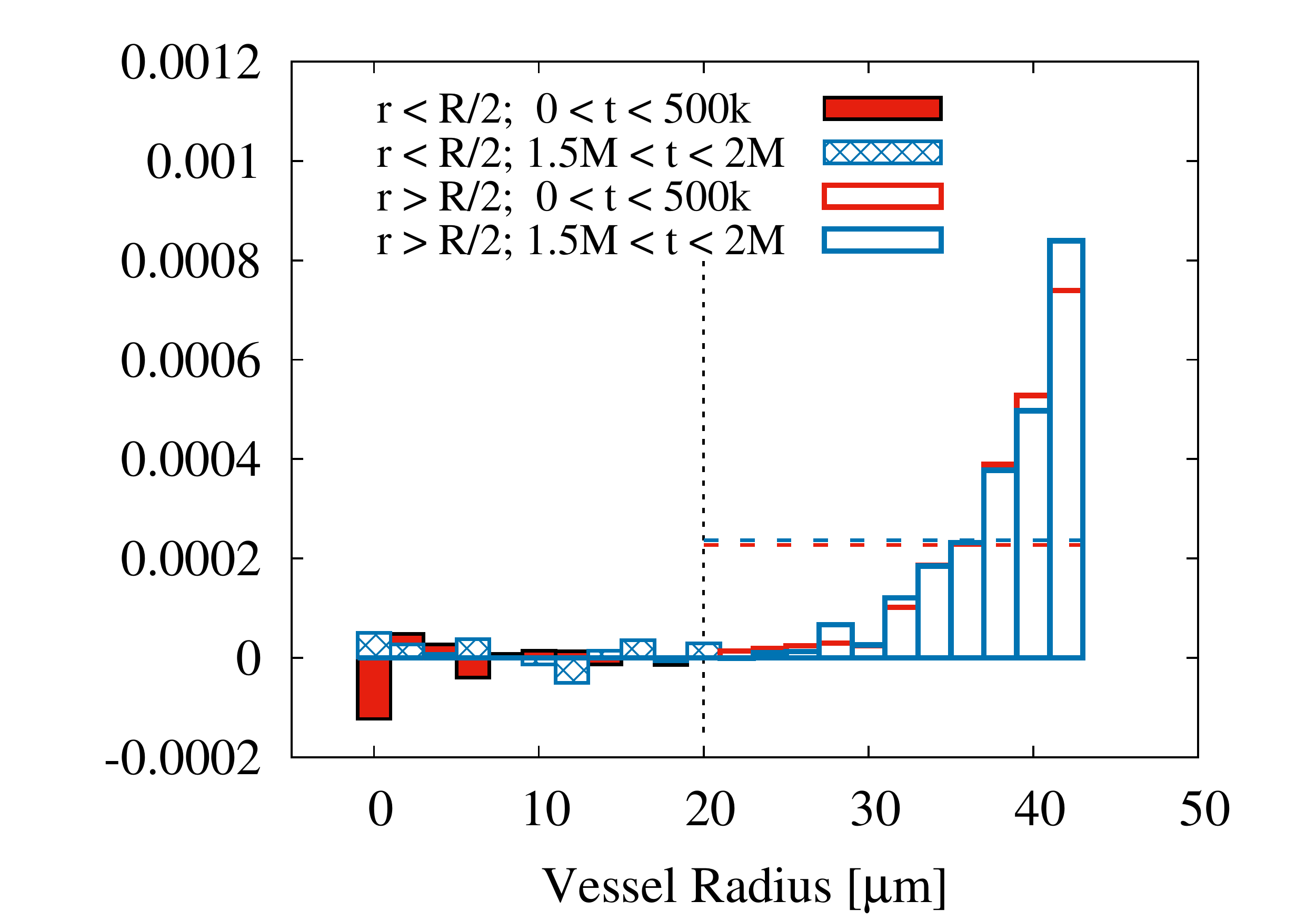}}
			\caption{\label{Rad_Vel_R_10} Time evolution of the radial velocity along the vessel radius ($R=40 \mu m$). In all cases, the spheres located at $r < R/2$ do not experience any significant sensible velocity towards the outer part of the vessel; on the other hand, the spheres at $r> R/2$ are characterized by an appreciable \textit{margination} behavior, which tends to decrease as time goes on (except for the case corresponding to $N_{SPH} = 3 \% N_{RBC}$). From the panel, it is apparent that the case corresponding to $N_{SPH} = 0.5 \% N_{RBC}$ is characterized by the largest margination velocity, as well as by the largest decrease in the radial velocity as time proceeds. The vertical dotted line represents $r=R/2$.}
	\end{center}
\end{figure}

\begin{figure}
	\begin{center}
			\subfigure[$N_{SPH} = 0.5 \% N_{RBC}$]{\includegraphics[angle=-90,width=0.34\textwidth]{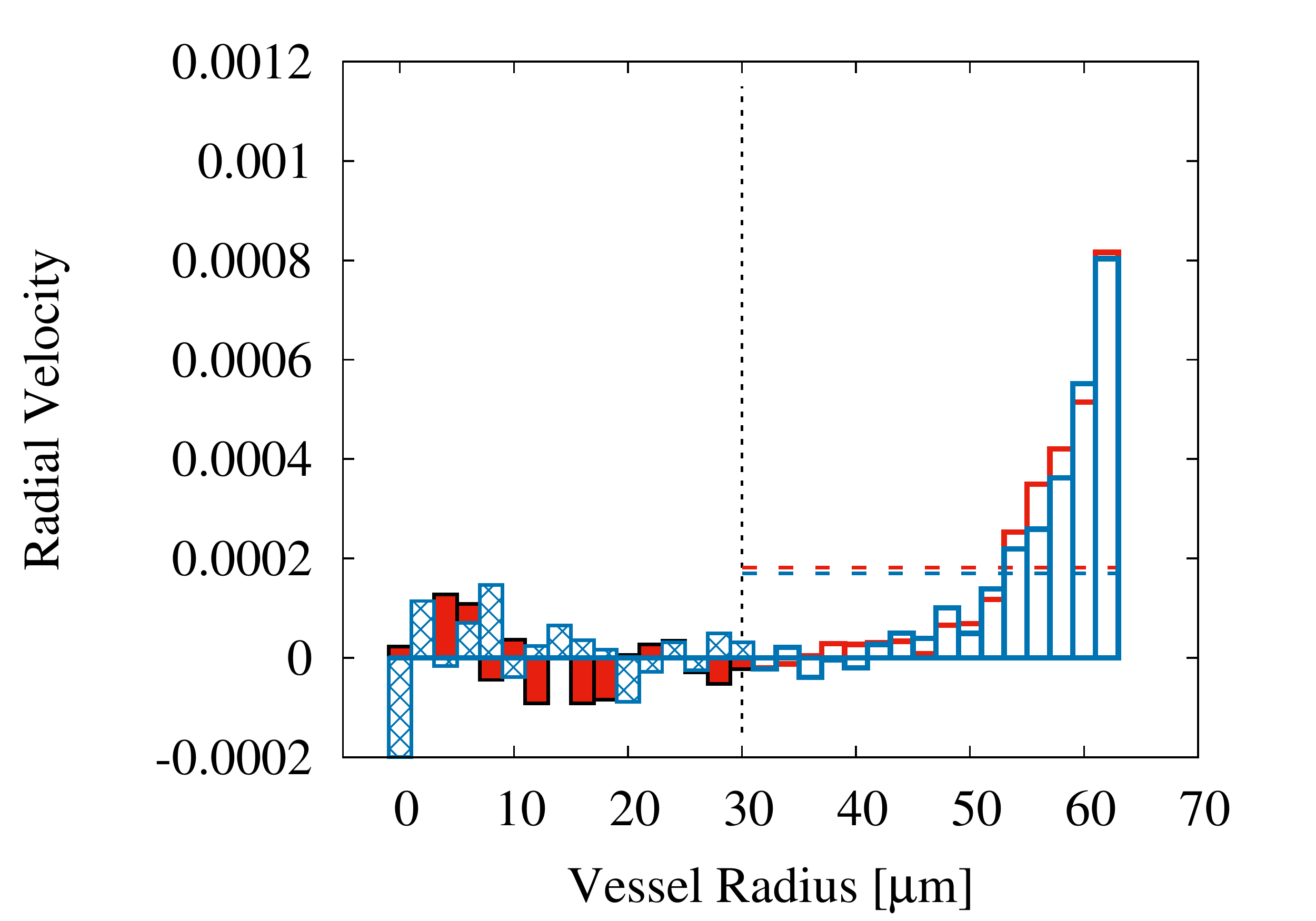}} \hspace{-0.5cm}
			\subfigure[$N_{SPH} = 1 \% N_{RBC}$]{\includegraphics[angle=-90,width=0.34\textwidth]{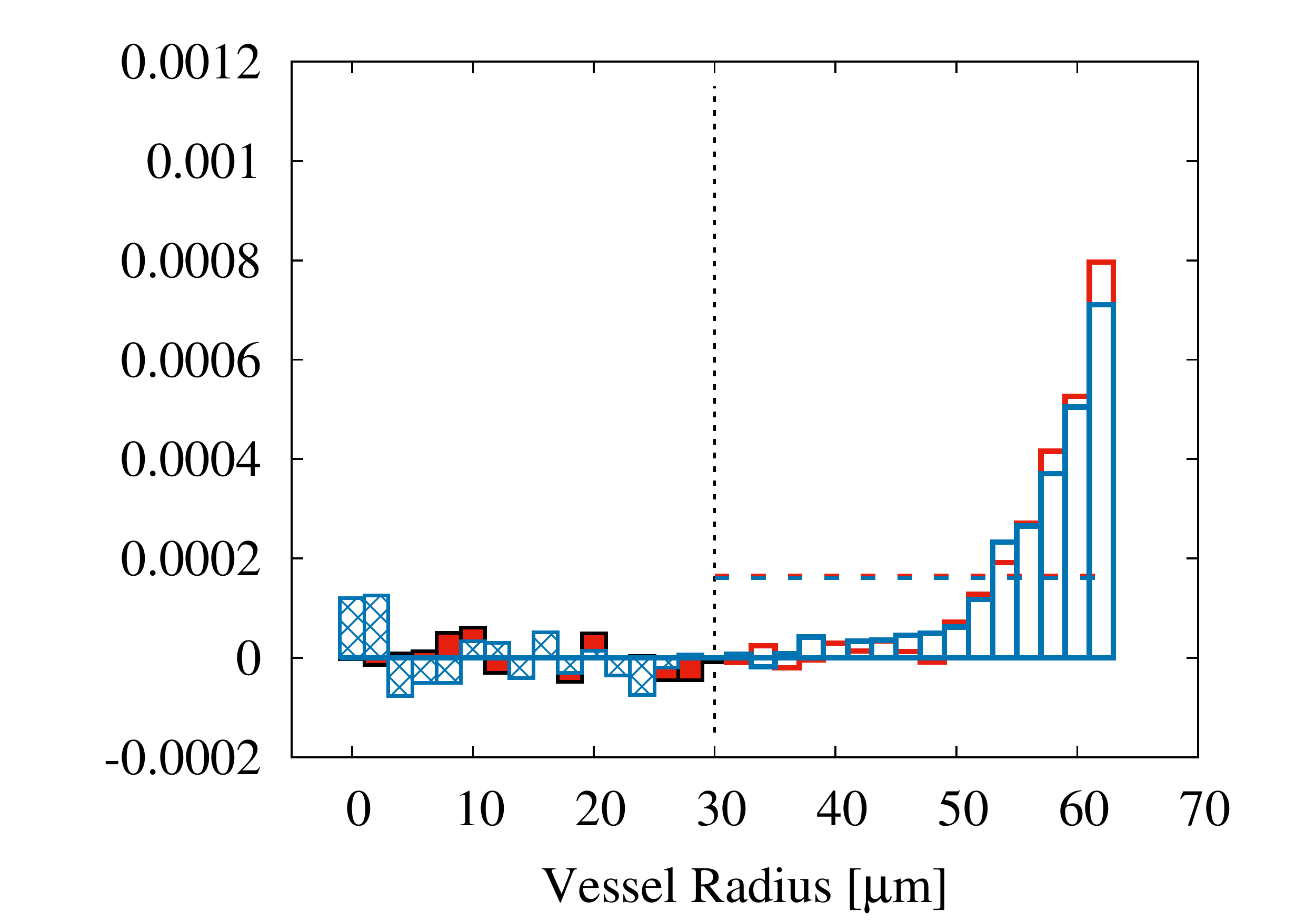}} \hspace{-0.5cm}
			\subfigure[$N_{SPH} = 3 \% N_{RBC}$]{\includegraphics[angle=-90,width=0.34\textwidth]{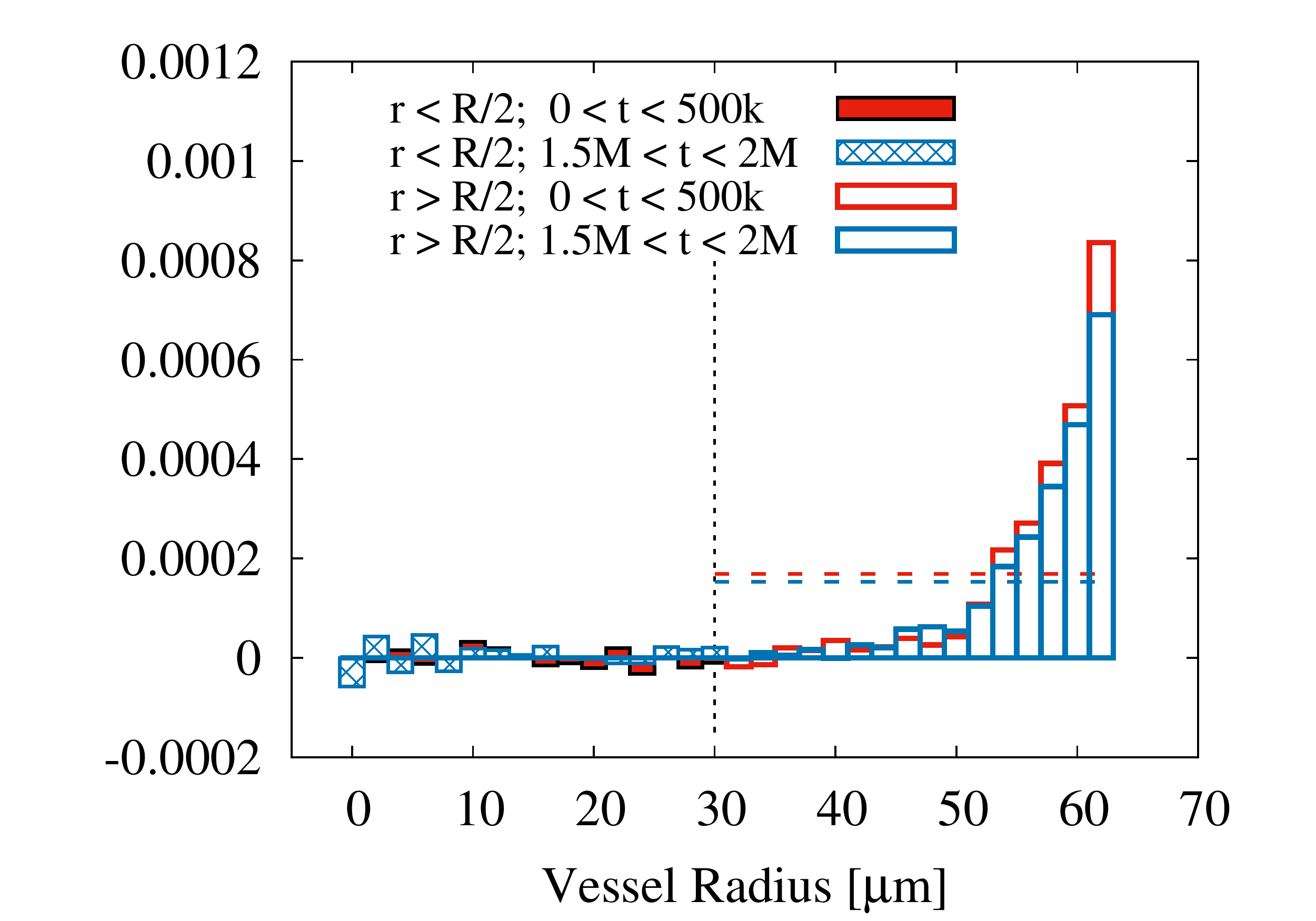}}
			\caption{\label{Rad_Vel_R_15} Time evolution of the radial velocity along the vessel radius ($R=60 \mu m$). As in Fig. \ref{Rad_Vel_R_10}, the spheres located at $r < R/2$ do not experience any significant velocity towards the outer part of the vessel; on the other hand, the spheres at $r> R/2$ are characterized by an appreciable \textit{margination} behavior, which tends to decrease as time goes on (except for the case corresponding to $N_{SPH} = 3 \% N_{RBC}$). From the panel, it is apparent that the case corresponding to $N_{SPH} = 0.5 \% N_{RBC}$ is characterized by the largest margination velocity, as well as by the largest decrease in the radial velocity as time proceeds. The vertical dotted line represents $r=R/2$.}
	\end{center}
\end{figure}

To confirm such a behavior, in Figures \ref{Rad_Vel_R_10} and \ref{Rad_Vel_R_15} report the distributions of radial velocity along the vessel radius for all the cases in Tab \ref{simcases} . \\
Interestingly, these Figures highlight that for both the channel dimensions, a positive radial velocity (i.e. pointing towards the outer walls of the vessel) is found for the particles located at $r>R/2$, providing further evidence of a \textit{margination} trend.
Such outward-velocity tends to decrease as the numerical simulation proceeds, reaching its maximum value in the first $350-500$k iterations, towards lower values, as the numerical simulations goes on. 
This is in line with expectations, as \textit{margination} is a transient process: considering an \textit{adequately long} computational time (theoretically, $t_{comp} \rightarrow \infty$), such an effect should disappear. \\
As to the particles located ad $r<R/2$, no evident margination is found: this too is a plausible behavior, which is due to the size of soft, micro-spheres, \cite{reasor2013determination} and to their initial random dispersion, \cite{reasor2013determination,karnadakis_communication}. Intuitively, this can be understood by considering the absence of any hydrodynamic force to push the spheres towards the outer part of the vessel, except for diffusion effects: such effects are characterized by very small velocities and the path of the spheres towards the vessel wall is hindered by the presence of RBC's, thus making such a migration an actual \textit{percolation} process.

\begin{figure}
\begin{center}
	\subfigure[]{\includegraphics[angle=-90,width=0.45\textwidth]{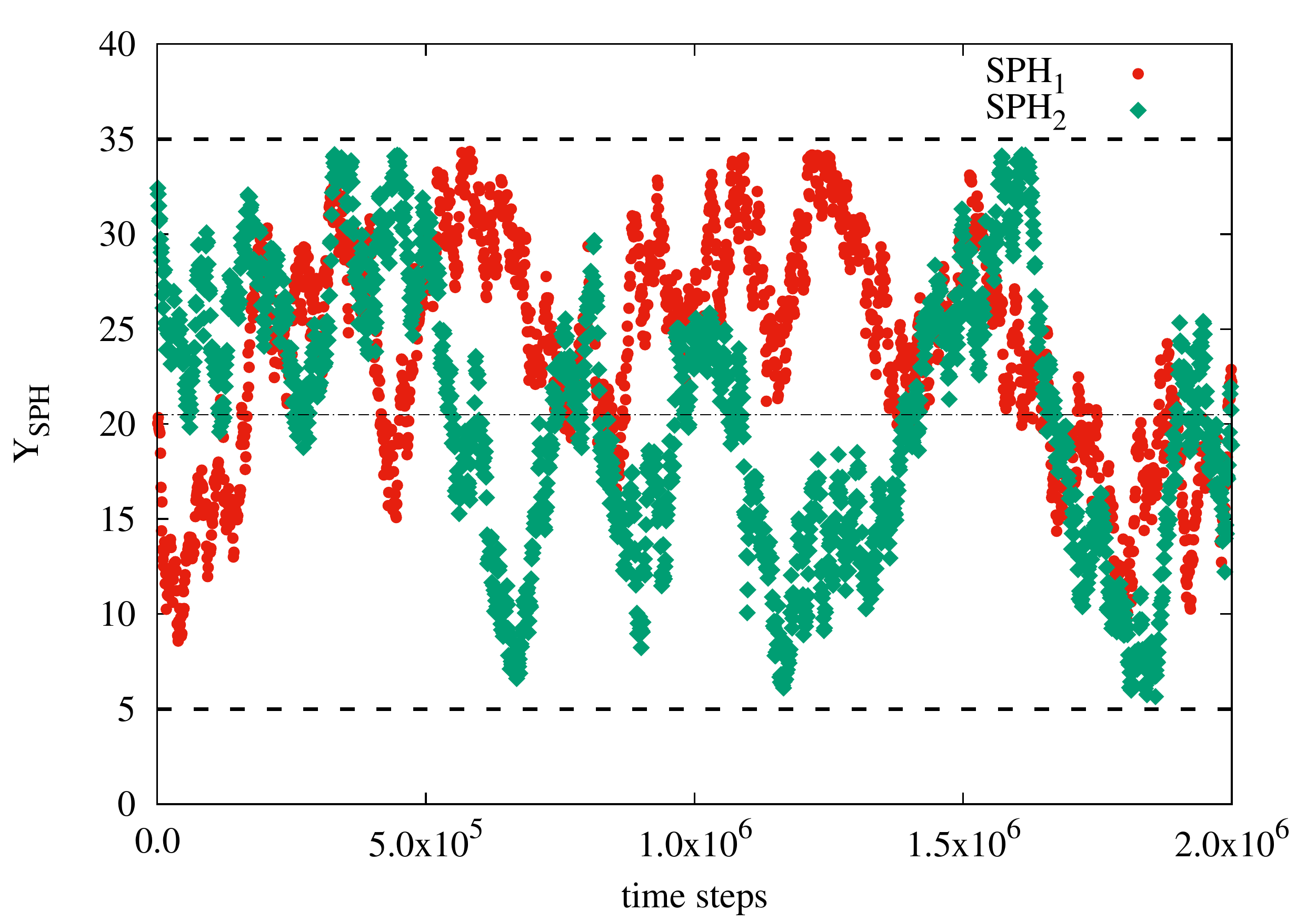}} \
	\subfigure[]{\includegraphics[angle=-90,width=0.45\textwidth]{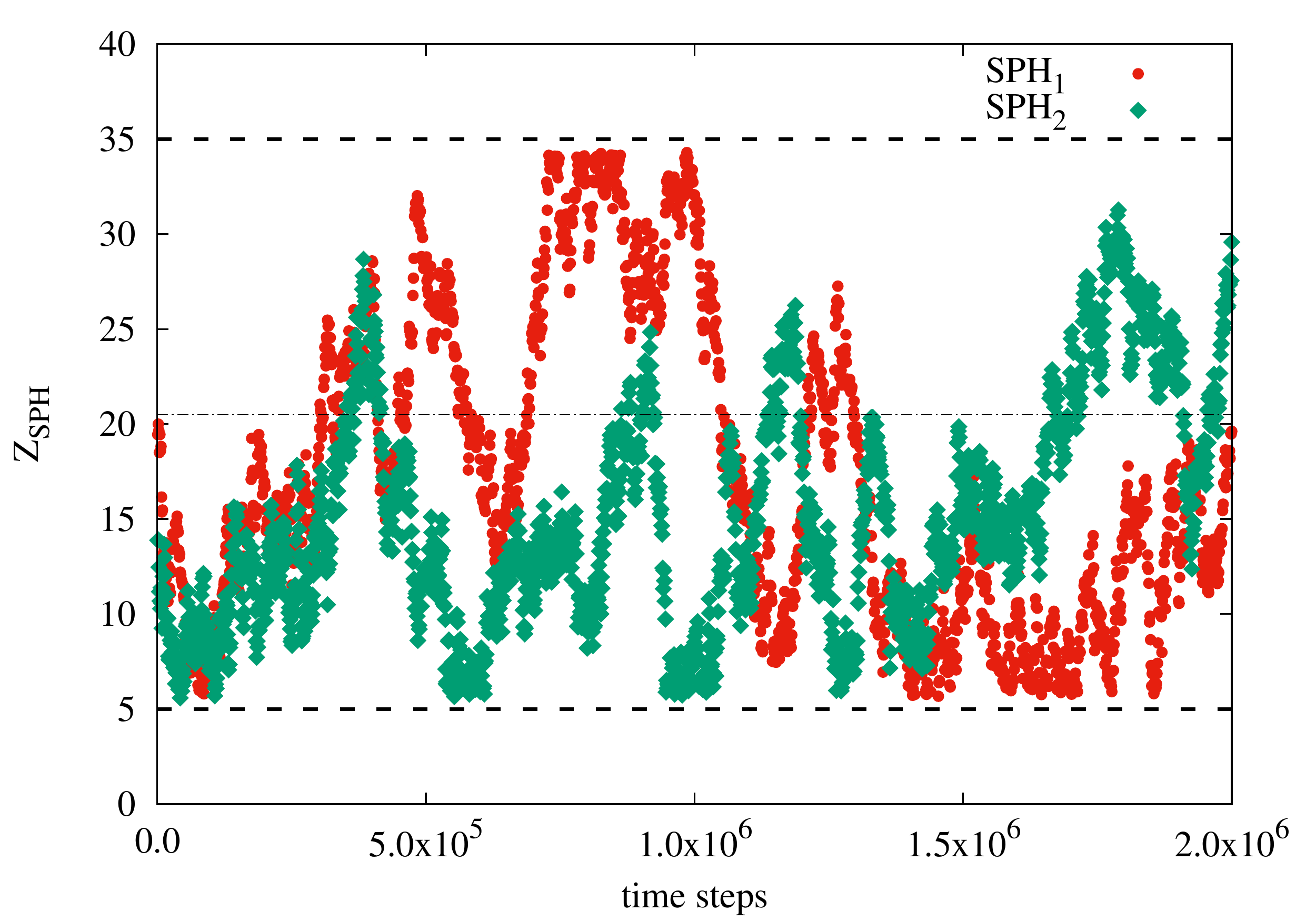}}
	\caption{\label{SPH_pos} Time evolution of the Y and Z locations of two selected spheres: SPH$_1$, which is initialized at the middle of the vessel, and SPH$_2$, initialized closed to the periphery. The panel highlights the trend of both spheres to \textit{marginate} towards the outer sections of the vessel. The dashed black lines represent the vessel walls.}
\end{center}
\end{figure}

To better highlight the margination behavior of transported spheres, in Figure \ref{SPH_pos} we report the evolution in time of the cross-section locations of two micro-spheres along the whole simulation.
SPH$_1$ is initialized in the middle of the vessel, while SPH$_2$ is located in the periphery at $t=0$. In this Figure, we refer to the case characterized by $R=60 \ \mu$m and $N_{SPH} = 3 \%$ $HCT$.
Both spheres are transported towards the outlet sections of the blood vessel throughout the numerical simulation: besides confirming the overall \textit{margination} behavior highlighted in Figs. \ref{Rad_Vel_R_10} and \ref{Rad_Vel_R_15}, it is possible to see that the actual path across a 1-radius length takes about $200-400$k iterations, supporting the percolation analogy for the margination process of transported micro-spheres.
Moreover, the panel shows that the sphere locations are reliably reconstructed by MUPHY algorithm at each time-step, providing a detailed time-history of spheres location without severely affecting the overall computational performance.

A final remark, in fact, should be addressed to the computational effort of the presented simulations with MUPHY.
The time-span of the simulations was chosen in order to ensure the steady state conditions in the flow rate of the plasma phase.
According to this criterion, the duration of the simulations is different for the various diameters of the vessel.
The reference Poiseuille flow simulations, in presence of just the plasma phase, were performed at an average $\sim 2$ Million Lattice Updates per Second (MLUPS).\\
In presence of RBC's, to simulate the F\aa hr\ae us--Lindqvist effect, the simulation time increased considerably, but it remained competitive, as related to the other methods in the literature, \cite{fedosov_predicting_2011,reasor2013determination}.
Considering, in fact, the largest simulated diameter, $D=160 \mu$m, corresponding to $150000$ fluid nodes, a $200000$ time step simulation has been performed. To ensure the hematocrit value of $HCT=40 \%$, $28457$ RBC's have been considered in the simulation.
The total computational time was $340000$ s, corresponding to $\sim 95$ h, for a total $\sim 0.1$ MLUPS. \\
In the presence of both RBC's and micro-spheres, since their number is much lower than that of erythrocytes, the computational effort is not sensibly affected by their presence.


\section{Methods}

\subsection{Blood plasma}

We model blood plasma by means of the Lattice Boltzmann (LB) method
\cite{SucciBook}. The LB method deals with the evolution
of the distribution function in discretized form over a cartesian
mesh, $f_{p}({\bf x},t)$, where ${\bf x}$ and $t$ are the spatial
and temporal coordinates and the subscript $p$ labels a set of discrete
speeds ${\bf c}_{p}$ connecting mesh points to mesh neighbors. The
evolution over a unit time step (i.e. $\Delta t = 1$) is given by
\begin{eqnarray}
f_{p}({\bf x}+{\bf c}_{p},t+ 1) & = & f_{p}^{*}({\bf x},t)\label{eq:streaming}
\end{eqnarray}
with $f_{p}^{*}({\bf x},t)$ being the post-collisional population,
\begin{eqnarray}
f_{p}^{*} & = & \left(1-\frac{1}{{\cal T}} \right) f_{p}+\frac{1}{{\cal T}} \ f_{p}^{eq}+\Delta f_{p}^{drag}\label{lbe_discrete}
\end{eqnarray}
Here ${\cal T}$ is a characteristic relaxation time and $f_{p}^{eq}$
the Maxwellian equilibrium expressed as a second-order low-Mach expansion
in the fluid velocity ${\bf u}$, $f_{p}^{eq}=w_{p}\rho\left[1+\frac{{\bf u}\cdot{\bf c}_{p}}{c^{2}}+\frac{({\bf u}\cdot{\bf c}_{p})^{2}-c^{2}u^{2}}{2c^{4}}\right]$.
The plasma kinematic viscosity $\nu$ relates to ${\cal T}$ via $\nu=c^{2}({\cal T}-1/2)$ where
$c$ is the plasma sound speed. We employ the D3Q19 lattice scheme,
where $c=1/\sqrt{3}$ , and $w_{p}$ is a set of normalized weights
with $p=0,...,18$, being equal to $w_{p}=1/3$ for the population
corresponding to the null discrete speed ${\bf c}_{0}=(0,0,0)$, $w_{p}=1/18$
for the ones connecting first mesh neighbors ${\bf c}_{1,...,6}=(\pm1,0,0),\,(0,\pm1,0),\,(0,0,\pm1)$,
and $w_{p}=1/36$ for second neighbors, ${\bf c}_{7,...,18}=(\pm1,\pm1,0),\,(\pm1,0,\pm1),\,(0,\pm1,\pm1)$. 

The term $\Delta f_{p}^{drag}$ accounts for the presence of suspended
particles that act as body forces on the plasma, with the following
expression
\begin{equation}
\Delta f_{p}^{drag}=h \ w_{p} \ \rho \left[\frac{{\bf G}\cdot{\bf c}_{p}}{c^{2}}+\frac{({\bf G}\cdot{\bf c}_{p})({\bf u}\cdot{\bf c}_{p})-c^{2}{\bf G}\cdot{\bf u}}{2c^{4}}\right]\label{eq:chen_force}
\end{equation}
where ${\bf G}$ is the local particle-fluid coupling. Eq. (\ref{eq:chen_force})
is a representation of the body force consistent with the second-order
Hermite expansion and BGK collisional kernel \cite{shan_kinetic_2006}.
Knowledge of the discrete populations $f_{p}$ allows to compute the
local plasma density $\rho$ and speed ${\bf u}$ as $\rho=\sum_{p}f_{p}$
and $\rho{\bf u}=\sum_{p}f_{p}{\bf c}_{p}+\frac{h}{2}\rho{\bf G}$,
ensuring second order space/time accuracy \cite{guo_discrete_2002}. 

\subsection{Red blood cells and suspended spheres}

We consider a suspension composed by ellipsoids and spheres. An oblate
ellipsoid approximates a RBC of mass $M$, position ${\bf R}_{i}$,
velocity ${\bf V}_{i}$, angular velocity ${\bm{\Omega}}_{i}$, and
instantaneous orientation given by the matrix 
\begin{equation}
{\bf Q}_{i}=\left(\begin{array}{ccc}
\hat{n}_{x,i} & \hat{t}_{x,i} & \hat{g}_{x,i}\\
\hat{n}_{y,i} & \hat{t}_{y,i} & \hat{g}_{y,i}\\
\hat{n}_{z,i} & \hat{t}_{z,i} & \hat{g}_{z,i}
\end{array}\right)
\end{equation}
where $\hat{{\bf n}}_{i}$, $\hat{{\bf t}}_{i}$, $\hat{{\bf g}}_{i}$
are orthogonal unit vectors, such that ${\bf Q}_{i}^{T}{\bf Q}_{i}={\bf 1}$.
The tensor of inertia, ${\bf I}_{i}$, is diagonal in the body frame
and transforms to the laboratory frame according to $\text{{\bf I}}'_{i}={\bf Q}_{i}\text{{\bf I}}_{i}\text{{\bf Q}}_{i}^{T}$.
Let us now introduce an auxiliary (``shape'')  function to account for the shape
and orientation of the suspended RBC. We choose the following expression
\cite{peskin_immersed_2002} 
\begin{equation}
\tilde{\delta}({\bf x},{\bf Q}_{i})\equiv\prod_{\alpha=x,y,z}\tilde{\delta}_{\alpha}[({\bf {\bf Q}}_{i}{\bf x})_{\alpha}]
\end{equation}
with
\begin{equation}
\tilde{\delta}_{\alpha}(y_{\alpha})\equiv\left\{ \begin{array}{cc}
\frac{1}{8}\left(5-4|y_{\alpha}/\xi_{\alpha}|-\sqrt{1+8|y_{\alpha}|/\xi_{\alpha}-16y_{\alpha}^{2}/\xi_{\alpha}^{2}}\right)\;\;\;\; & |y_{\alpha}/\xi_{\alpha}|\leq0.5\\
\frac{1}{8}\left(3-4|y_{\alpha}|/\xi_{\alpha}-\sqrt{-7+24|y_{\alpha}|/\xi_{\alpha}-16y_{\alpha}^{2}/\xi_{\alpha}^{2}}\right)\;\;\;\; & 0.5<|y_{\alpha}/\xi_{\alpha}|\leq1\\
0 & |y_{\alpha}|/\xi_{\alpha}>1
\end{array}\right.
\end{equation}
and $\xi_{\alpha}$ being a set of three integers, one for each cartesian
component $\alpha=x,y,z$, representing the ellipsoidal radii in the
three principal directions. The shape function is normalized as \cite{peskin_immersed_2002},
 $\sum_{{\bf x}}\tilde{\delta}({\bf x}-{\bf J})=1$ for any continuous displacement ${\bf J}$, and obeys 
 $\sum_{{\bf x}}(x_{\alpha}-T_{\alpha})\partial_{\beta}\tilde{\delta}({\bf x}-{\bf T})=-\delta_{\alpha\beta}$.
The translational motion of the particle-fluid systems is given by
the following exchange kernel
\begin{equation}
\boldsymbol{\phi}({\bf {\bf x}},{\bm{\Gamma}}_{i})=-\gamma_{T} \ \tilde{\delta} \ ({\bf x}-{\bf R}_{i},{\bf Q}_{i})\left[{\bf V}_{i}-{\bf u}({\bf x})\right]\label{eq:pheno_force}
\end{equation}
where $\gamma_{T}$ is a translational coupling coefficient.

The rotational response is obtained by decomposing the deformation tensor
in terms of purely elongational and rotational terms $\boldsymbol{\partial}{\bf u}=\boldsymbol{e}+\boldsymbol{\rho}$
where ${\bf e}=\frac{1}{2}(\boldsymbol{\partial}{\bf u}+\boldsymbol{\partial}{\bf u}^{T})$
is the symmetric rate of strain tensor, related to the dissipative
character of the flow, and $\boldsymbol{\rho}=\frac{1}{2}(\boldsymbol{\partial}{\bf u}-\boldsymbol{\partial}{\bf u}^{T})$
is the antisymmetric vorticity tensor, which bears the conservative
component of the flow and is related to the vorticity vector $\boldsymbol{\omega}=\boldsymbol{\partial}\times{\bf u}$
\cite{leal2007advanced}. The rotational component of the deformation
tensor induces tumbling motion, while the rotational and elongational
one give rise to the vesicular, tank treading motion. Consequently,
suspended particles experience the coupling between the body motion
and the fluid vorticity, that we represent by the following rotational
kernel 
\begin{equation}
\boldsymbol{\tau}^{A}({\bf {\bf x}},{\bm{\Gamma}_{i}})=-\gamma_{R} \ \tilde{\delta}({\bf x}-{\bf R}_{i},{\bf Q}_{i})\left[{\bm{\Omega}}_{i}-{\bf \boldsymbol{\omega}}({\bf x})\right]=-\gamma_{R} \ \tilde{\delta}_{i}\left({\bm{\Omega}}_{i}-{\bf \boldsymbol{\omega}}\right)\label{eq:pheno_torque}
\end{equation}
where $\gamma_{R}$ is a rotational coupling coefficient and the superscript
$A$ stands for antisymmetric. This term depends on the particle shape
and instantaneous orientation. The elongational component of the flow
contributes to the orientational torque for bodies with ellipsoidal
symmetry, being zero for spherical solutes \cite{rioual2004analytical}.
By defining the stress vector ${\bf t}^{\boldsymbol{\sigma}}=\boldsymbol{\sigma}\cdot\hat{{\bf n}}$,
where $\hat{{\bf n}}$ is the outward normal to the surface of a suspended
particles, we replace the surface normal with the vector spanning
over the entire volume of the diffused particle, $\hat{{\bf n}}=\boldsymbol{\partial}\tilde{\delta}/|\boldsymbol{\partial}\tilde{\delta}|$.
The associated torque is represented in analogy with the torque acting
on macroscopic bodies \cite{leal_advanced_2007}, by the kernel 
\begin{equation}
\boldsymbol{\tau}^{S}({\bf {\bf x}},{\bm{\Gamma}}_{i})= \alpha \ \tilde{\delta}_{i} \ {\bf t}^{\boldsymbol{\sigma}}\times({\bf x}-{\bf R}_{i})\label{eq:pheno_torque_diss}
\end{equation}
where $\alpha$ is a parameter to be fixed and the superscript $S$
is mnemonic for the symmetric contribution of the flow. 

The hydrodynamic force and torque acting on the suspended particles
are obtained via integration over the particle spatial extension,
written as 
\begin{eqnarray}
{\bf F}_{i} & = & \sum_{{\bf x}}\boldsymbol{\phi}({\bf {\bf x}},{\bm{\Gamma}}_{i})=-\gamma_{T}({\bf V}_{i}-\tilde{{\bf u}}_{i})\\
{\bf T}_{i}^{A} & = & \sum_{{\bf x}}\boldsymbol{\tau}^{A}({\bf {\bf x}},{\bm{\Gamma}}_{i})=-\gamma_{R}({\bm{\Omega}}_{i}-\tilde{\boldsymbol{\omega}}_{i})\\
{\bf T}_{i}^{S} & = & \sum_{{\bf x}}\boldsymbol{\tau}^{S}({\bf {\bf x}},{\bm{\Gamma}}_{i})
\end{eqnarray}
where $\tilde{{\bf u}}_{i}\equiv\sum_{{\bf x}}\tilde{\delta}_{i}{\bf u}$
and $\tilde{\boldsymbol{\omega}}_{i}\equiv\sum_{{\bf x}}\tilde{\delta}_{i}\boldsymbol{\omega}$.
The action of forces and torques on the fluid populations is given
by
\[
{\bf G}=-\sum_{i}\left\{ {\bf F}_{i}\tilde{\delta}_{i}+\frac{1}{2}{\bf T}_{i}\times\boldsymbol{\partial}\tilde{\delta}_{i}\right\} 
\]

\subsection{Viscosity contrast}

Suspended particles typically carry an internal fluid composed of
hemoglobin proteins, as in the case of RBC. The highly viscous inner
fluid is taken into account by a local enhancement of the LB fluid
viscosity within the particle shape according to the following BGK
relaxation time
\begin{equation}
{\cal T}(x)={\cal T}_{0}+\Delta h\sum_{i}\tilde{\theta}_{i}\label{eq:enhance-viscosity}
\end{equation}
where ${\cal T}_{0}$ corresponds to the viscosity of pure plasma,
$\Delta$ is viscosity enhancement factor, and $\tilde{\theta}_{i}$
is a smooth version of the particle characteristic function. The latter
is represented as $\tilde{\theta}_{i}=1-(1-\tilde{\delta_{i}})^{\kappa}$
where the parameter $\kappa$ governs the smooth transition between
the inner and outer fluids, chosen as $\kappa=20$. By choosing $\nu_{0}=1/6$
and $\Delta=2$, the ratio between inner ($\tilde{\theta}_{i}\sim1$)
and outer ($\tilde{\theta}_{i}\sim0$) viscosities is equal to $5$.

\subsection{Excluded Volume Interactions}

RBC-RBC, sphere-sphere and RBC-sphere repulsive forces due soft-core
mechanical forces and torques are taken into account via the Gay-Berne
(GB) potential \cite{gay_modification_1981}, by introducing an orientation-dependent
repulsive interaction derived from the Lennard-Jones potential ($\phi_{LJ}(R_{ij})=\epsilon[(\sigma/R_{ij})^{12}-(\sigma/R_{ij})^{6}]$,
where $\epsilon$ is the energy scale and $\sigma$ the ``contact''
length scale). 

Given the principal axes $(a_{i,1},a_{i,2},a_{i,3})$ of the $i$-th
particle, the shape associated to the excluded volume interactions
is constructed according to the shape matrix $S_{i}=\mbox{{diag}}(a_{i,1},a_{i,2},a_{i,3})$
and the transformed matrix ${\bf A}_{i}={\bf Q}_{i} \ {\bf S}_{i}^{2} \ {\bf Q}_{i}^{T}$
in the laboratory frame. Two particles $i,j$ at distance ${\bf R}_{ij}$
have an exclusion distance $\sigma_{ij}$ that depends on the mutual
distance, shape and mutual orientation, written as
\begin{eqnarray}
\sigma_{ij} & = & \frac{1}{\sqrt{\phi_{ij}}}\label{eq:sigmagb}\\
\phi_{ij} & = & \frac{1}{2}\hat{{\bf R}}_{ij}\cdot{\bf H}_{ij}^{-1}\cdot\hat{{\bf R}}_{ij}\label{eq:phigb}
\end{eqnarray}
where ${\bf H}_{ij}\equiv{\bf A}_{i}+{\bf A}_{j}$. A purely repulsive
exclusion potential is given by \cite{gay_modification_1981,allen_expressions_2006}
\begin{equation}
u_{ij}=\left\{ \begin{array}{lc}
4\epsilon_{0}(\rho_{ij}^{-12}-\rho_{ij}^{-6})+\epsilon_{0} & \qquad\rho_{ij}^{6}\leq2\\
0 & \qquad\rho_{ij}^{6}>2
\end{array}\right.\label{eq:gayberne}
\end{equation}
with
\begin{equation}
\rho_{ij}=\frac{R_{ij}-\sigma_{ij}+\sigma_{ij}^{min}}{\sigma_{ij}^{min}}
\end{equation}
where $\epsilon_{0}$ is the energy scale and $\sigma_{ij}^{min}$
is a constant, both parameters being independent on the ellipsoidal
mutual orientation and distance. By considering the minimum particle
dimension $a_{i}^{min}=\min(a_{i,1},a_{i,2},a_{i,3})$ then 
\begin{equation}
\sigma_{ij}^{min}=\sqrt{2\left[\left(a_{i}^{min}\right)^{2}+\left(a_{j}^{min}\right)^{2}\right]}
\end{equation}

For a confining environment, we associate to each wall mesh point
a spherical particle that acts as a repulsive center for RBCs and
spheres. The wall-particle characteristic distance $\sigma_{w,RBC}^{min}$
is chosen to be half the mesh spacing $\frac{\Delta x}{2}$.

\section{Conclusions}

In this work we have shown the capability of the MUPHY code, based on the LBPD framework, to reproduce the main aspects related to blood flow circulation and transport of micro-particles.\\
The F\aa hr\ae us--Lindqvist effect of blood viscosity variation with vessel diameter was accurately reproduced in presence of a physiological hematocrit value ($HCT = 40 \%$) and for a wide range of diameters.
A lack of agreement was found for very small capillaries ($D<40 \mu$m), for which the deformability of RBC's and vessel walls become dominant in the overall flow. \\
The presence of micro-scale, non-deformable, spheres has been simulated, as well: the effect of sphere margination towards the outer of the blood vessel was detected for different values of sphere concentrations and vessel diameter, by analyzing their spacial distribution in the channel cross section at the end of the simulation and their radial velocity and the evolution in time of cross-section locations during the whole numerical simulation ($t = 2 \times 10^6$ time steps). \\
The results highlight the reliability and versatility of MUPHY to accurately reproduce complex transport phenomena in biological systems, in presence of consistent hematocrit values and accounting for the presence of transported micro-particles, as well. \\
As a future development, RBC and wall deformability will be accounted for, in order to increase the accuracy of the simulations in the smaller capillaries; moreover, micro- and nano-scale objects of different shapes will be implemented, to simulate the delivery and target of transported media within the blood circulation.



\section*{Acknowledgements}
SS wishes to acknowledge financial support
from the European Research Council under the European
Union's Horizon 2020 Framework Programme (No. FP/2014-
2020)/ERC Grant Agreement No. 739964 (COPMAT). \\
G.F. wishes to thank the support by the Italian Ministry Program PRIN, grant n. 20154EHYW9. \\
The numerical simulations were performed on \textit{Zeus} HPC facility, 
at the University of Naples ``Parthenope'';  
\textit{Zeus} HPC has been realized through the Italian Government Grant 
PAC01$\_$00119 ``MITO - Informazioni Multimediali per Oggetti Territoriali'', 
with Prof. E. Jannelli as the Scientific Responsible. \\
Fruitful and inspiring discussions with Prof. George Em Karniadakis are kindly acknowledged.
%
%
\section*{Author contributions statement}
%
G.F. performed the numerical simulations; G.F., S.M., P.D. and S.S. analyzed the data. All authors reviewed the manuscript. 
%
%
%
%
%
%

\end{document}